\documentclass[number,sort&compress,12pt]{elsarticle}
\usepackage{graphicx}
\usepackage{amssymb}
\usepackage{amsmath}
\usepackage{xcolor}
\usepackage{graphicx}
\usepackage{subcaption}
\usepackage{hyperref}

\journal{Physica A}

\begin{document}

\begin{frontmatter}

\title{Still life in a classic Blume-Capel model: pseudo-transitions in a spin-1 diamond chain}

\author[UPJS]{Elham Shahhosseini Shahrabadi}
\ead{elham.shhoseini@gmail.com}

\author[UPJS]{Majid Moradi}
\ead{majidofficial@gmail.com}

\author[UPJS]{Jozef Stre\v{c}ka\corref{cor}}
\ead{jozef.strecka@upjs.sk}

\address[UPJS]{Department of Theoretical Physics and Astrophysics, Faculty of Science, \\
P. J. \v{S}af\'{a}rik University, Park Angelinum 9, 040 01 Ko\v{s}ice, Slovak Republic}
	
\cortext[cor]{Corresponding author.}

\begin{abstract}
We exactly investigate the ground-state, magnetic, and thermodynamic properties of a spin-1 Blume-Capel diamond chain in a magnetic field by means of the transfer-matrix method. After establishing the complete ground-state phase diagram and characterizing each ground-state spin configuration, we examine the finite-temperature behavior in the vicinity of selected phase boundaries and triple points. It is demonstrated that an extremely small energy gap between a nondegenerate ground state and competing macroscopically degenerate low-lying excited states gives rise to entropically-driven pseudo-transitions. These pseudo-transitions manifest themselves through abrupt but continuous changes in the magnetization and entropy resembling discontinuous jumps, while the magnetic susceptibility and specific heat exhibit exceptionally sharp yet finite peaks resembling divergences. Our results provide the exact evidence that pseudo-transitions can also emerge in the classical spin-1 Blume-Capel model and thereby extend the class of one-dimensional spin systems known to display pseudo-transitions.
\end{abstract}

\end{frontmatter}

\section{Introduction}

A collective mechanism leading to magnetic ordering and phase transitions associated with spontaneous symmetry breaking is fundamentally described by discrete lattice-statistical spin models \cite{niss}.
This line of research was triggered by the paradigmatic spin-$1/2$ Ising model, whose exact solution for a one-dimensional spin chain initially led to the erroneous conclusion that spontaneous long-range magnetic order is absent at any non-zero temperature regardless of the system dimensionality \cite{lenz, ising1925}. This misconception was decisively overturned by Onsager's exact solution of the spin-$1/2$ Ising model on a two-dimensional square lattice \cite{onsager}, which provided the rigorous proof of a finite-temperature phase transition between the ferromagnetically ordered and disordered paramagnetic phases.  

According to the fundamental non-existence theorems of van Hove \cite{vanhove}, Cuesta and S\'anchez \cite{cuesta}, finite-temperature phase transitions are strictly forbidden in one-dimensional systems with only short-range and non-singular interactions. Despite these limitations, a particular class of one-dimensional spin systems can exhibit anomalous magnetic and thermodynamic features that closely resemble true phase transitions \cite{landau1984, toledano1987}. These phenomena originate from a pronounced thermally-activated crossover between the nondegenerate ground state and massively degenerate low-lying excited states and are commonly referred to as pseudo-transitions \cite{Souza2018,rojas2019,ORojas2020}. Unlike genuine phase transitions, however, the one-dimensional systems do not exhibit at these pseudo-transitions true mathematical singularities in their thermodynamic potentials. Instead, first-order derivatives of the Gibbs free energy such as the magnetization and entropy undergo exceptionally steep yet continuous changes that closely mimic discontinuous jumps. Similarly, second-order response functions including the magnetic susceptibility and specific heat exhibit sharp but finite peaks rather than true power-law divergences.

Pseudo-transitions have been reported in a broad variety of one-dimensional lattice-statistical models with diverse microscopic interactions and lattice topologies. The first exactly solvable example was a coupled spin-electron double-tetrahedral chain composed of localized Ising spins and mobile electrons \cite{PRE2015}. This was followed by several spin-$1/2$ Ising-Heisenberg models with the geometry of ladder \cite{ladder2016}, tube \cite{tube2016}, diamond chain \cite{CARVALHO2019}, and double-tetrahedral chain \cite{RojasStrecka2020}. Pseudo-transitions were subsequently identified also in conceptually much simpler spin-1/2 Ising diamond and tetrahedral chains \cite{dia2020,strecka2020Ising}, a spin-1/2 Ising ladder with trimer rungs \cite{kro21,Sznajd2022,Yin24}, a mixed spin-(1/2, 1) Ising nanowire \cite{PIMENTA2022}, and a spin-1/2 Ising Toblerone tube \cite{chapman2024}. More recently, the concept has been further extended to a broad class of decorated spin-$1/2$ Ising chains in a magnetic field \cite{Yin2024}, a sawtooth chain \cite{Yin2025}, a site-decorated chain \cite{Yin2026}, and a twin-diamond chain \cite{rojas2026}. Notably, nearly all of these studies concern either pure spin-$1/2$ or mixed spin-$(1/2,1)$ systems that can be exactly mapped onto effective spin-$1/2$ Ising models.

Compared to this, pseudo-transitions in purely spin-1 Ising systems and their extensions have received considerably less attention. Among the notable extensions of the spin-1 Ising model is the spin-1 Blume–Capel model introduced independently by Blume \cite{blume1966} and Capel \cite{capel1966}, which supplements the nearest-neighbor exchange interaction with a uniaxial single-ion anisotropy. The one-dimensional spin-1 Blume-Capel chain was solved exactly by means of the transfer-matrix method, yet no evidence of pseudo-transitions was found \cite{lines1979, Chatterjee1984}. Similarly, the exact solution for the spin-1 Blume-Capel sawtooth chain in a magnetic field revealed no signatures of pseudocritical behavior \cite{sawtoothBC2016}. By contrast, pseudo-transitions have recently been reported for the closely related  spin-1 Blume–Emery–Griffiths sawtooth chain in a magnetic field, where the inclusion of a biquadratic interaction fundamentally alters the low-temperature behavior and gives rise to quasi-ordering phenomena  \cite{zvyagin2026, zvyagin2026quasi}.  These findings naturally raise the question of whether pseudo-transitions can emerge in the spin-1 Blume-Capel systems without invoking additional interaction terms beyond the bilinear exchange interaction and uniaxial single-ion anisotropy. 

To address this issue, we investigate in the present work the spin-1 Blume-Capel diamond chain in a magnetic field. This model represents a reasonable candidate for exhibiting pseudo-transitions as its spin-1/2 Ising counterpart is known to display pronounced pseudocritical behavior \cite{dia2020,strecka2020Ising}. Beyond its theoretical relevance, the studies of diamond spin chains are also of considerable experimental interest because this exotic one-dimensional magnetic structure is realized in several low-dimensional magnetic compounds. For instance, the natural mineral azurite is widely recognized as a prototype of a spin-$1/2$ diamond-chain material \cite{KIKUCHI2003, Kikuchi2005, kikuchi2005oxford}, whereas spin-1 diamond-chain architectures can be realized in coordination polymers based on $\text{Ni}^{2+}$ ions \cite{guillou2002, Ni2026}. Although the magnetic properties of transition-metal compounds are generally captured by the Heisenberg rather than the Ising model, $\text{Ni}^{2+}$ ions may experience a relatively strong uniaxial single-ion anisotropy of easy-axis type. Under such conditions, the spin-1 Blume-Capel diamond chain provides a simplified yet physically well-motivated model capable of capturing the essential magnetic properties of $\mathrm{Ni}^{2+}$-based diamond-chain compounds at least in the strong-anisotropy regime. 

The remainder of this paper is organized as follows. In Sec.~2, we introduce the spin-1 Blume-Capel diamond chain and briefly summarize its exact transfer-matrix solution. Section~3 presents the main results including the ground-state phase diagram, the characterization of the individual ground states, and a detailed analysis of the pseudo-transition phenomenon through temperature dependencies of the magnetization, magnetic susceptibility, entropy, and specific heat in several representative parameter regimes. Finally, the main findings and their implications for pseudo-transitions in one-dimensional spin-1 systems are summarized in Sec.~4.

\section{Model and method}
\label{model}

\begin{figure}
\begin{center}
\includegraphics[width=0.7\columnwidth]{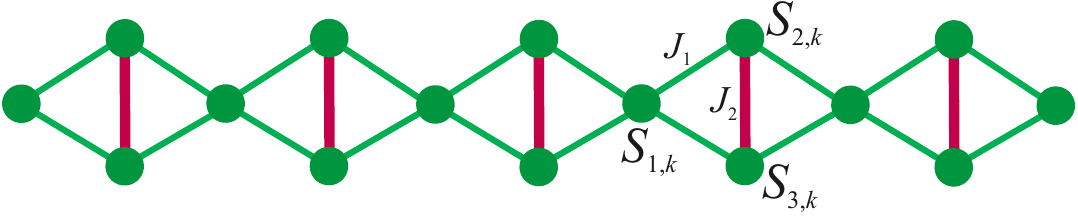}
\end{center}
\vspace{-0.5cm} 
\caption{Schematic illustration of the spin-1 Blume-Capel diamond chain.}
\label{fig1}
\end{figure}

We consider a spin-1 diamond chain described by the Blume-Capel model, which accounts for the uniaxial single-ion anisotropy and an external magnetic field. The magnetic structure of the investigated system is illustrated in Fig.~\ref{fig1}. The Hamiltonian of the spin-1 Blume-Capel diamond chain in the presence of external magnetic field is given by:
\begin{align}
	\mathcal{H} &=
	J_1 \sum_{k=1}^{N}
	\left(S_{2,k}+S_{3,k}\right)
	\left(S_{1,k}+S_{1,k+1}\right)
	+J_2 \sum_{k=1}^{N} S_{2,k}S_{3,k}
	\nonumber \\
	&+D\sum_{k=1}^{N}
	\left(S_{1,k}^2+S_{2,k}^2+S_{3,k}^2\right)
	-h\sum_{k=1}^{N}
	\left(S_{1,k}+S_{2,k}+S_{3,k}\right),
	\label{htot}
\end{align}
where $S_{\alpha,k} \in \{0, \pm1\}$ denote the spin-1 variables assigned to the three sites $\alpha=1,2,3$ of the $k$-th unit cell, the parameter $J_1>0$ represents the antiferromagnetic interaction between the nodal and decorating spins, while the parameter $J_2>0$ denotes the antiferromagnetic interaction within the decorating spin pairs. Notably, the competition between these two antiferromagnetic couplings gives rise to geometric frustration and may stabilize macroscopically degenerate ground states. Furthermore, the parameters $D$ and $h$ stand for the uniaxial single-ion anisotropy and the Zeeman energy associated with the presence of an external magnetic field, respectively. A positive single-ion anisotropy $D>0$ favors the nonmagnetic state $S_{\alpha,k}=0$, while a negative single-ion anisotropy $D<0$ stabilizes two magnetic states $S_{\alpha,k}=\pm 1$. For simplicity, we impose the periodic boundary condition $S_{1,N+1} \equiv S_{1,1}$, which will significantly simplify an exact treatment within the transfer-matrix approach.

For further analysis, it is convenient to rewrite the total Hamiltonian~\eqref{htot} as a sum over cell Hamiltonians:
\begin{equation}
	\mathcal{H} = \sum_{k=1}^N \mathcal{H}_k,
	\label{hk}
\end{equation}
where the cell Hamiltonian $\mathcal{H}_k$ contains all interaction terms belonging to the $k$th diamond unit cell:
\begin{equation}
	\mathcal{H}_k = \mathcal{H}'_k + \frac{D}{2} \left(S_{1,k}^2 + S_{1,k+1}^2\right)
	- \frac{h}{2} \left(S_{1,k} + S_{1,k+1}\right),
\end{equation}
with
\begin{equation}
	\mathcal{H}'_k = J_1 (S_{2,k} + S_{3,k})(S_{1,k} + S_{1,k+1})
	+ J_2 S_{2,k} S_{3,k}
	+ D (S_{2,k}^2 + S_{3,k}^2)
	- h (S_{2,k} + S_{3,k}).
\end{equation}
This decomposition separates the total Hamiltonian into individual cell contributions, which enables at the level of the the partition function to trace out spin degrees of freedom of the decorating spin pairs $S_{2,k}$ and $S_{3,k}$ before tracing out spin degrees of freedom of the nodal spins $S_{1,k}$:
\begin{align}
	\mathcal{Z} = \sum_{\{S_{1,k}\}} \prod_{k=1}^{N} \sum_{S_{2,k}} \sum_{S_{3,k}} 
	\exp \left(-\beta \mathcal{H}_k\right).
	\label{pfd}
\end{align}
Here, $\sum_{\{S_{1,k}\}}$ denotes the summation over all possible configurations of the nodal spins $S_{1,k}$, $\beta = 1/(k_{\rm B} T)$, $k_{\rm B}$ is the Boltzmann constant, and $T$ is the absolute temperature.

To proceed with the transfer-matrix method, the expression behind the product symbol in Eq. (\ref{pfd}) can be identified with the transfer matrix $\mathbf{T}$ whose elements are explicitly defined as:
\begin{equation}
	\mathbf{T}_{(S_{1,k}, S_{1,k+1})} = \exp \Big[ -\frac{\beta D}{2} (S_{1,k}^2 + S_{1,k+1}^2) + \frac{\beta h}{2} (S_{1,k} + S_{1,k+1}) \Big] 
	\sum_{S_{2,k}, S_{3,k}} \!\!\! \exp (-\beta \mathcal{H}'_k),
	\label{tm1}
\end{equation}
where the Boltzmann factor $\sum_{S_{2,k}, S_{3,k}} \exp (-\beta \mathcal{H}'_k)$ gains the following explicit form after performing the two summations over spin degrees of freedom of the decorating spins from $k$th diamond unit cell:
\begin{align}
	\smash[b]{\sum_{S_{2,k}, S_{3,k}}} \!\!\! \exp (-\beta \mathcal{H}'_k) &=
	1 + 2\exp \left[-\beta (2D-J_2)\right]  \\
	&+ 4\exp \left(-\beta D\right) \cosh \Big\{ \beta \big[ J_1 (S_{1,k} + S_{1,k+1}) - h \big] \Big\} \nonumber\\
	&+ 2\exp \left[-\beta (2D+J_2)\right] \cosh \Big\{ 2\beta \big[ J_1 (S_{1,k} + S_{1,k+1}) - h \big] \Big\}. 
	\label{tm2}
\end{align}
After performing a subsequent summation over spin degrees of freedom of all nodal spins except the first one, the partition function can be expressed within the transfer-matrix method as a trace of the $N$th power of the transfer matrix $\mathbf{T}$:
\begin{align}
	\mathcal{Z} = \sum_{\{S_{1,k}\}} \prod_{k=1}^{N} \mathbf{T}_{(S_{1,k}, S_{1,k+1})} 
	       = \sum_{S_{1,1}=0,\pm 1} \mathbf{T}^N_{(S_{1,1}, S_{1,1})} = \mbox{Tr} \, \mathbf{T}^N.
	\label{pff}
\end{align}
The evaluation of the partition function therefore reduces to determining the eigenvalues of the transfer matrix $\mathbf{T}$:
\begin{align}
	\mathbf {T} = \left( \begin{array}{lll}
	T_{(1,1)} & ~T_{(1,0)} & ~T_{(1,-1)} \\
	T_{(0,1)} & ~T_{(0,0)} & ~T_{(0,-1)} \\
	T_{(-1,1)} & ~T_{(-1,0)} & ~T_{(-1,-1)}
	\end{array}
	\right),
\end{align}
whose individual elements are defined by Eqs. (\ref{tm1})-(\ref{tm2})
The eigenvalues of this matrix are obtained as the roots of the corresponding cubic characteristic equation:
\begin{align}
	\lambda^3 - a\lambda^2 + b\lambda + c = 0,
	\label{CE}
\end{align}
where the coefficients $a$, $b$, and $c$ are defined as:
\begin{align}
	a &= T_{(0,0)} + T_{(-1,-1)} + T_{(1,1)}, \nonumber\\
	b &= T_{(1,1)} T_{(0,0)} + T_{(1,1)} T_{(-1,1)} + T_{(0,0)} T_{(-1,-1)} - T^2_{(0,-1)} - T^2_{(1,0)} - T^2_{(1,-1)},
	\nonumber\\
	c &= T_{(1,1)} T^2_{(0,-1)} + T_{(0,0)} T^2_{(1,-1)} + T_{(-1,-1)} T^2_{(1,0)} \nonumber\\
	 & - T_{(1,1)} T_{(0,0)} T_{(-1,-1)} - 2 \: T_{(1,0)} T_{(0,-1)} T_{(-1,1)}.
\end{align}
The explicit expressions for the three eigenvalues of the transfer matrix read:
\begin{equation}
	\lambda_n = \frac{a}{3} + 2 \operatorname{sign}(q) \sqrt{p} \cos \Big[ \frac{1}{3} (\varphi + 2n\pi) \Big], \quad n = 1, 2, 3,
	\label{RCE}
\end{equation}
where
\begin{align}
	p &= \Big(\frac{a}{3}\Big)^2 - \frac{b}{3}, \qquad
	q = \Big(\frac{a}{3}\Big)^3 - \frac{ab}{6} - \frac{c}{2}, \nonumber \\
	\varphi &= \arctan \left( \frac{\sqrt{p^3 - q^2}}{q} \right), \quad \varphi \in \left(-\frac{\pi}{2}, \frac{\pi}{2}\right).
\end{align}
The partition function ${\cal Z}$ given by Eq.~(\ref{pff}) is thus obtained from the eigenvalues $\lambda_1$, $\lambda_2$, and $\lambda_3$ of the transfer matrix according to:
\begin{align}
	\mathcal{Z} = \mathrm{Tr}\, \mathbf{T}^N
	= \lambda_1^N + \lambda_2^N + \lambda_3^N.
	\label{pffi}
\end{align}

In the thermodynamic limit $N \to \infty$, the Gibbs free energy per unit cell is solely determined by the dominant eigenvalue of the transfer matrix $\lambda_{0} = \max\left\{\lambda_1, \lambda_2, \lambda_3\right\}$:
\begin{align}
	g = -\frac{1}{\beta} \lim_{N \to \infty} \frac{1}{N} \ln {\cal Z} = -\frac{1}{\beta} \ln \lambda_{0},
	\label{frebege}
\end{align}
From the Gibbs free energy, the total magnetization $M$ per spin, the magnetic susceptibility $\chi$ per spin, 
the entropy $S$ per unit cell, and the specific heat $C$ per unit cell can be easily calculated using the following standard relations:
\begin{align}
	 M &= - \frac{1}{3} \frac{\partial g}{\partial h} = \frac{1}{3 \beta \lambda_{0}} \frac{\partial \lambda_{0}}{\partial h}, \nonumber \\
	\chi &= \frac{\partial M}{\partial h} =
	\frac{1}{\beta} \left[\frac{1}{\lambda_{0}} \frac{\partial^2 \lambda_{0}}{\partial h^2} -
	\frac{1}{\lambda_{0}^2} \left(\frac{\partial \lambda_{0}}{\partial h}\right)^2 \right], \nonumber \\
	S &= k_{\rm{B}} \beta^2  \frac{\partial g}{\partial \beta}
	= k_{\rm{B}} \left( \ln \lambda_{0} - \frac{\beta}{\lambda_{0}} \frac{\partial \lambda_{0}}{\partial \beta}\right) , \nonumber \\
	C &= - \beta  \frac{\partial S}{\partial \beta}
	= k_{\rm{B}} \beta^2   \left[\frac{1}{\lambda_{0}} \frac{\partial^2 \lambda_{0}}{\partial \beta^2} -
	\frac{1}{\lambda_{0}^2} \left(\frac{\partial \lambda_{0}}{\partial \beta}\right)^2 \right].
	\end{align}

\section{Results and discussion}
\label{result}

\begin{figure}
	\begin{center}
		\includegraphics[width=\textwidth]{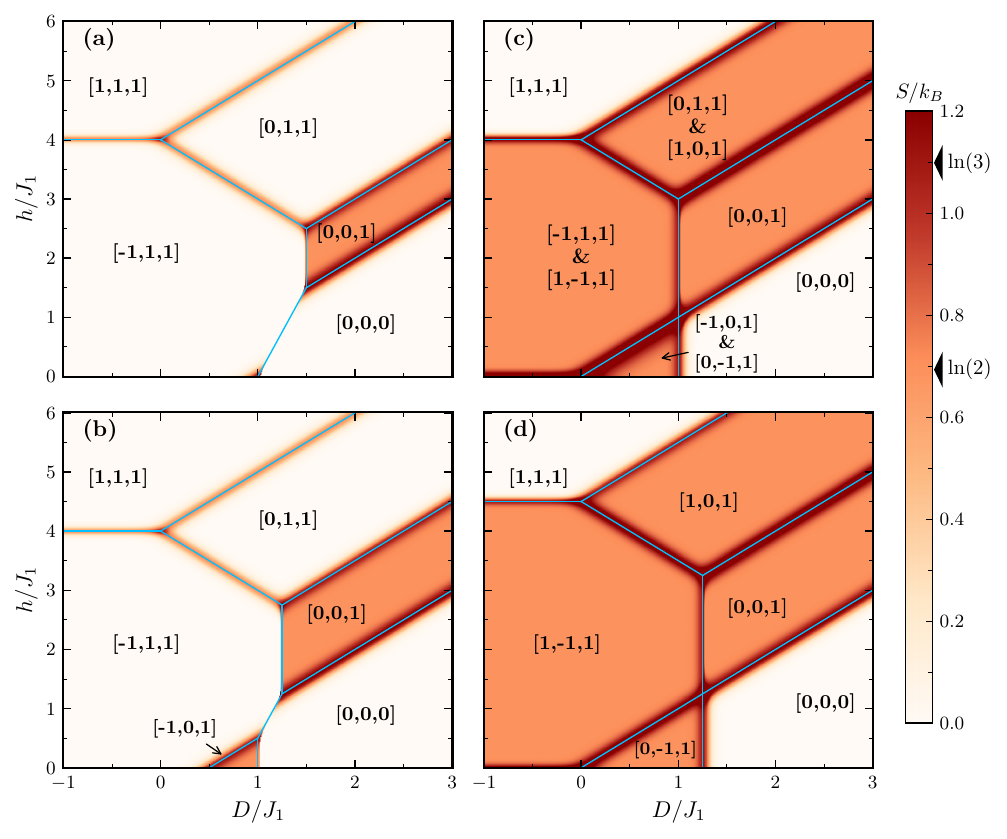}
	\end{center}
	\vspace{-0.5cm}
	\caption{Ground-state phase diagram of the spin-1 Blume–Capel diamond chain in the $D/J_1-h/J_1$ plane for four representative values of the interaction ratio: (a) $J_2/J_1 = 1.0$, (b) $J_2/J_1 = 1.5$, (c) $J_2/J_1 = 2.0$, and (d) $J_2/J_1 = 2.5$. Light blue lines represent ground-state phase boundaries given by Eq. (\ref{boundary}), while the color map represents a density plot of the entropy per unit cell $S/k_{\rm B}$ calculated at sufficiently low temperature $k_{\rm B}T/J_1 = 0.05$. The parameter regions with an orange color correspond to macroscopically degenerate ground states with the residual entropy $S/k_{\rm B}=\ln(2)$ (see an arrow on a color scale), while a darker color traces a higher entropy characteristic for coexistence lines and triple points.}
	\label{fig2}
\end{figure}

In this section, we present the exact results obtained for the spin-1 Blume–Capel diamond chain in an external magnetic field. To reduce the total number of independent model parameters, the coupling constant $J_1$ between the decorating and nodal spins is adopted as an energy unit when defining the dimensionless parameters $J_2/J_1$, $D/J_1$, $h/J_1$, and $k_{\rm B}T/J_1$ measuring a relative strength of the coupling constants, uniaxial single-ion anisotropy, magnetic field, and temperature, respectively.    

\subsection{Ground-state phase diagram}

The ground-state spin arrangement is determined by a subtle interplay between the competing antiferromagnetic interactions, the uniaxial single-ion anisotropy, and the magnetic field. Throughout this section, each ground states is specified by the spin configuration $[S_1,S_2,S_3]$ of the nodal and decorating spins from the unit cell. The spin-1 Blume–Capel diamond chain hosts in total nine distinct ground states $[0,0,0]$, $[-1,+1,+1]$, $[0,+1,+1]$, $[+1,+1,+1]$, $[0,0,+1]$, $[-1,0,+1]$, $[0,-1,+1]$, $[+1,-1,+1]$, [+1,0,+1], whose corresponding energies per unit cell $\mathcal{E}_{[S_1, S_2, S_3]}$ are given by:  
\begin{align}
	\mathcal E_{[0,0,0]} &= 0, \nonumber \\
	\mathcal E_{[-1,+1,+1]} &= -4J_1 + J_2 + 3D - h, \nonumber \\
	\mathcal E_{[0,+1,+1]} &= J_2 + 2D - 2h, \nonumber \\
	\mathcal E_{[+1,+1,+1]} &= 4J_1 + J_2 + 3D - 3h, \nonumber \\
	\mathcal E_{[0,0,+1]} &= D-h, \nonumber \\
	\mathcal E_{[-1,0,+1]} &= -2J_1 + 2D, \nonumber \\
	\mathcal E_{[0,-1,+1]} &= -J_2 + 2D, \nonumber \\
	\mathcal E_{[+1,-1,+1]} &= -J_2 + 3D - h, \nonumber \\
	\mathcal E_{[+1,0,+1]} 	&= 2J_1 + 2D - 2h .
	\label{EE}
\end{align}
Owing to the spatial symmetry of the decorating spins under the interchange $S_2 \leftrightarrow S_3$, every ground-state spin configuration with unequal states of the decorating spins $S_2 \neq S_3$ has an equivalent counterpart obtained by exchanging the two decorating spins, i.e. from $[S_1,S_2,S_3]$ to $[S_1,S_3,S_2]$. The five ground states $[0,0,+1]$, $[-1,0,+1]$, $[+1,0,+1]$, $[+1,-1,+1]$, and $[0,-1,+1]$ are accordingly macroscopically degenerate with the residual entropy $S = k_{\rm B} \ln(2)$ arising from a two-fold degeneracy of each unit cell. 

Let us now shortly describe the individual ground states emerging in the ground-state phase diagram, which is shown in Fig.~\ref{fig2} in $D/J_1-h/J_1$ plane for four representative values of the interaction ratio $J_2/J_1$. The nonmagnetic ground state $[0,0,0]$ dominates the parameter region at low magnetic fields and sufficiently strong positive single-ion anisotropies of the easy-plane type $D>0$ irrespective of the interaction ratio $J_2/J_1$. However, the character of remaining ground states depends basically on the relative strength of the competing antiferromagnetic interactions $J_2/J_1$. If considering first the parameter region with the dominant coupling constant between the nodal and decorating spins $J_2/J_1 \leq 1$ as exemplified in Fig.~\ref{fig2}(a), the ferrimagnetic ground state $[-1,+1,+1]$ is stabilized at low magnetic fields and sufficiently weak single-ion anisotropy. At moderate values of the external magnetic field, one additionally encounters the ground states $[0,0,+1]$ and $[0,+1,+1]$ characterized by the nonmagnetic nodal spins and either partially or fully polarized decorating spins, respectively. When the interaction ratio is selected from the interval $1<J_2/J_1<2$, the ground-state phase diagram additionally acquires the macroscopically degenerate antiferromagnetic ground state $[-1,0,+1]$, which appears in between the ferrimagnetic $[-1,+1,+1]$ and nonmagnetic $[0,0,0]$ ground states as illustrated in Fig.~\ref{fig2}(b) for the interaction ratio $J_2/J_1 = 1.5$. The ground-state spin arrangements changes dramatically at the special value of the interaction ratio $J_2/J_1 = 2$, where the three previously described ground states coexist with three other macroscopically degenerate ground states, see Fig.~\ref{fig2}(c). In particular, the ferrimagnetic phase $[-1,+1,+1]$ coexists with the macroscopically degenerate ferrimagnetic phase $[+1,-1,+1]$, the antiferromagnetic phase $[-1,0,+1]$ coexists with another macroscopically degenerate antiferromagnetic phase $[0,-1,+1]$, and the phase $[0,+1,+1]$ coexists with the macroscopically degenerate phase $[+1,0,+1]$. Notably, a strong spin frustration originating from the dominant interaction between the decorating spins $J_2/J_1 > 2$ stabilizes the latter three ground states $[+1,-1,+1]$, $[0,-1,+1]$, $[+1,0,+1]$ and fully suppresses their counterparts $[-1,+1,+1]$, $[-1,0,+1]$, $[0,+1,+1]$, respectively. The ground-state phase diagram shown in Fig.~\ref{fig2}(d) for the interaction ratio $J_2/J_1 = 2.5$ consequently contains just the four macroscopically degenerate ground states $[+1,-1,+1]$, $[0,-1,+1]$, $[0,0,+1]$, $[+1,0,+1]$ in addition to the nonmagnetic ground state $[0,0,0]$ and the fully polarized ferromagnetic phase $[+1,+1,+1]$. For completeness,  we list the analytical expressions for the ground-state phase boundaries determining a coexistence of two phases, which are obtained by equating their corresponding unit cell energies:
\begin{align}
	\mathcal E_{[-1,+1,+1]} &= \mathcal E_{[0,0,+1]} \implies \frac{D}{J_1} = 2 - \frac{J_2}{2J_1} , \nonumber \\
	\mathcal E_{[-1,+1,+1]} &= \mathcal E_{[-1,0,+1]} \implies \frac{h}{J_1} = \frac{D}{J_1} + \frac{J_2}{J_1} - 2, \nonumber \\
	\mathcal E_{[-1,+1,+1]} &= \mathcal E_{[0,0,0]} \implies \frac{h}{J_1} = 3\frac{D}{J_1} + \frac{J_2}{J_1} - 4, \nonumber \\
	\mathcal E_{[0,0,+1]} &= \mathcal E_{[0,0,0]} \implies \frac{h}{J_1} = \frac{D}{J_1}, \nonumber \\
	\mathcal E_{[-1,0,+1]} &= \mathcal E_{[0,0,0]} \implies  \frac{D}{J_1} = 1,\nonumber \\
	\mathcal E_{[0,-1,+1]} &= \mathcal E_{[0,0,0]} \implies  \frac{D}{J_1} = \frac{J_2}{2J_1}.
	\label{boundary}
\end{align}
In the following, we demonstrate that the vicinity of selected ground-state phase boundaries and triple coexistence points provides favorable conditions for the emergence of pseudo-transitions. 
These anomalous finite-temperature features originate from the thermal activation of macroscopically degenerate low-lying excited states, whose entropic contribution may overcome a very small energy gap with respect to a competing nondegenerate ground state.

\subsection{Pseudo-transitions of the spin-1 Blume-Capel diamond chain}

\begin{figure}[t]
	\begin{center}
		\includegraphics[width=0.95\textwidth]{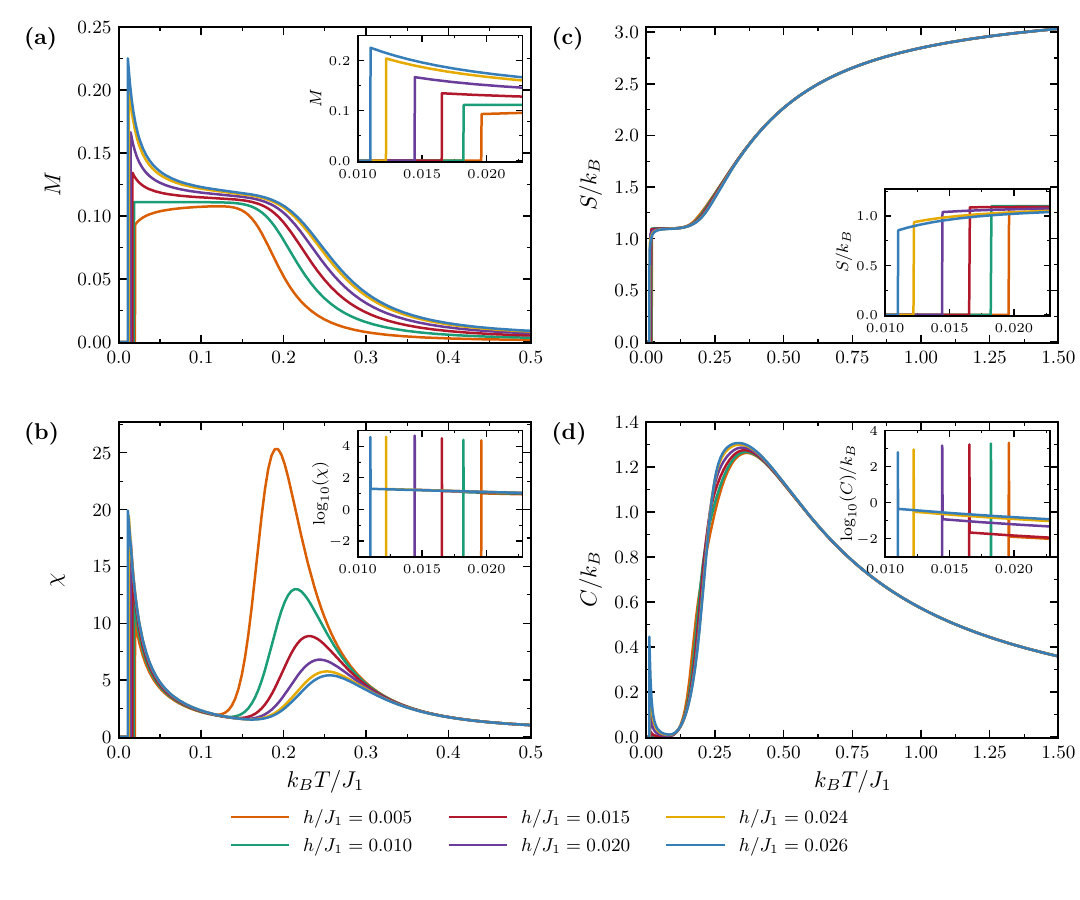}
	\end{center}
	\vspace{-0.5cm}
	\caption{Temperature dependence of the magnetization $M$ per spin (a), the susceptibility $\chi$ per spin (b), the entropy $S/k_{\rm B}$ per unit cell (c), and the specific heat $C/k_{\rm B}$ per unit cell (d) for the spin-1 Blume-Capel diamond chain with the interaction ratio $J_2/J_1=1.0$, the single-ion anisotropy $D/J_1=1.01$, and magnetic fields $h/J_1 \in [0.005, 0.026]$. Insets show the pseudo-transition region in an enhanced scale.}
	\label{fig3}
\end{figure}

Before discussing the most interesting numerical results, let us briefly clarify the physical mechanism driving the pseudo-transitions. According to the van Hove theorem and its generalizations \cite{vanhove, cuesta}, true finite-temperature phase transitions accompanied by spontaneous symmetry breaking are in one-dimensional systems governed by short-range interactions and non-singular potential strictly forbidden. Nevertheless, exceptionally sharp crossovers accompanied by abrupt yet continuous changes resembling discontinuities and/or extremely sharp but finite peaks resembling divergences may still occur when a one-dimensional system is driven to a unique ground state separated by only a very small energy gap from competing massively degenerate excited states. This condition is naturally satisfied sufficiently close to a ground-state boundary between the nondegenerate and competing macroscopically degenerate states, which allows a substantial redistribution of the dominant Boltzmann weights even by a slight increase in temperature as the entropic gain associated with the highly degenerate excited states compensates for their small energy excess. This entropy-driven crossover gives rise to a pseudo-transition, which closely mimics a genuine phase transition despite the absence of any nonanalyticity in the thermodynamic potentials.

First, we illustrate in Fig.~\ref{fig3} the pseudo-transition phenomenon for the spin-1 Blume-Capel diamond chain with the parameter set $J_2/J_1 = 1.0$ and $D/J_1 = 1.01$, which drives the system toward the unique nonmagnetic ground state $[0,0,0]$ positioned in close proximity to the ground-state phase boundary with the ferrimagnetic state $[-1,+1,+1]$ (see Eq.~\eqref{boundary}). Although the competing ferrimagnetic state $[-1,+1,+1]$ is itself nondegenerate, the nonmagnetic ground state $[0,0,0]$ coexists at zero field not only with the ferrimagnetic state $[-1,+1,+1]$ but also with the macroscopically degenerate states $[-1,0,+1]$. All these competing excited states are separated from the ground state by an extremely small energy gap, which enables their substantial thermal population at ultra-low temperatures as manifested by a narrow reddish region in Fig.~\ref{fig2}(a) developing above the ground-state phase boundary between the $[0,0,0]$ and $[-1,+1,+1]$ ground states even at the relatively low temperature $k_{\rm B} T/J_1 = 0.05$.

The temperature dependencies of the magnetization and magnetic susceptibility consequently display highly anomalous behavior at sufficiently low magnetic fields. The magnetization remains zero only in a relatively narrow range of temperatures and then exhibits an abrupt yet continuous thermally-activated rise around the pseudocritical temperature as highlighted in the inset of Fig.~\ref{fig3}(a). This anomalous feature progressively shifts toward lower temperatures upon increasing the magnetic field $h/J_1$ reflecting the gradual destabilization of the nonmagnetic ground state $[0,0,0]$. The sharp peak of the magnetization is finally followed by a two-step thermal disordering process with an intermediate plateau emergent around the moderate temperatures $k_{\rm B} T/J_1 \approx 0.1$. The two distinct magnetization changes are also reflected in a striking two-peak temperature dependencies of the magnetic susceptibility depicted in Fig.~\ref{fig3}(b). Specifically, the susceptibility displays an extremely sharp low-temperature peak in a narrow temperature interval around the pseudo-critical temperature (shown on a logarithmic scale in the inset) in addition to a much broader round high-temperature maximum originating from ordinary thermal excitations.

The entropy and specific heat provide even more compelling evidence for the pseudo-transition. As shown in  Fig.~\ref{fig3}(c), the entropy undergoes at pseudocritical temperature a steep but continuous increase from zero to an intermediate plateau near $S/k_{\rm B} = \ln(3) \approx 1.10$, which is successively followed by a gradual rise toward its high-temperature limit. This pronounced entropy gain reflects the thermal population of low-lying macroscopically degenerate excited states, whose large configurational degeneracy overcomes their small energy excess relative to the nondegenerate ground state. This marked entropy profile is consistent with a pronounced double-peak temperature dependence of the specific heat displayed in Fig.~\ref{fig3}(d). The pseudo-transition produces an exceptionally sharp yet finite peak of the specific heat at a low temperature, which is well separated from the broader high-temperature maximum originating from conventional thermal excitations involving higher-energy states.

\begin{figure}[t]
	\begin{center}
		\includegraphics[width=0.95\textwidth]{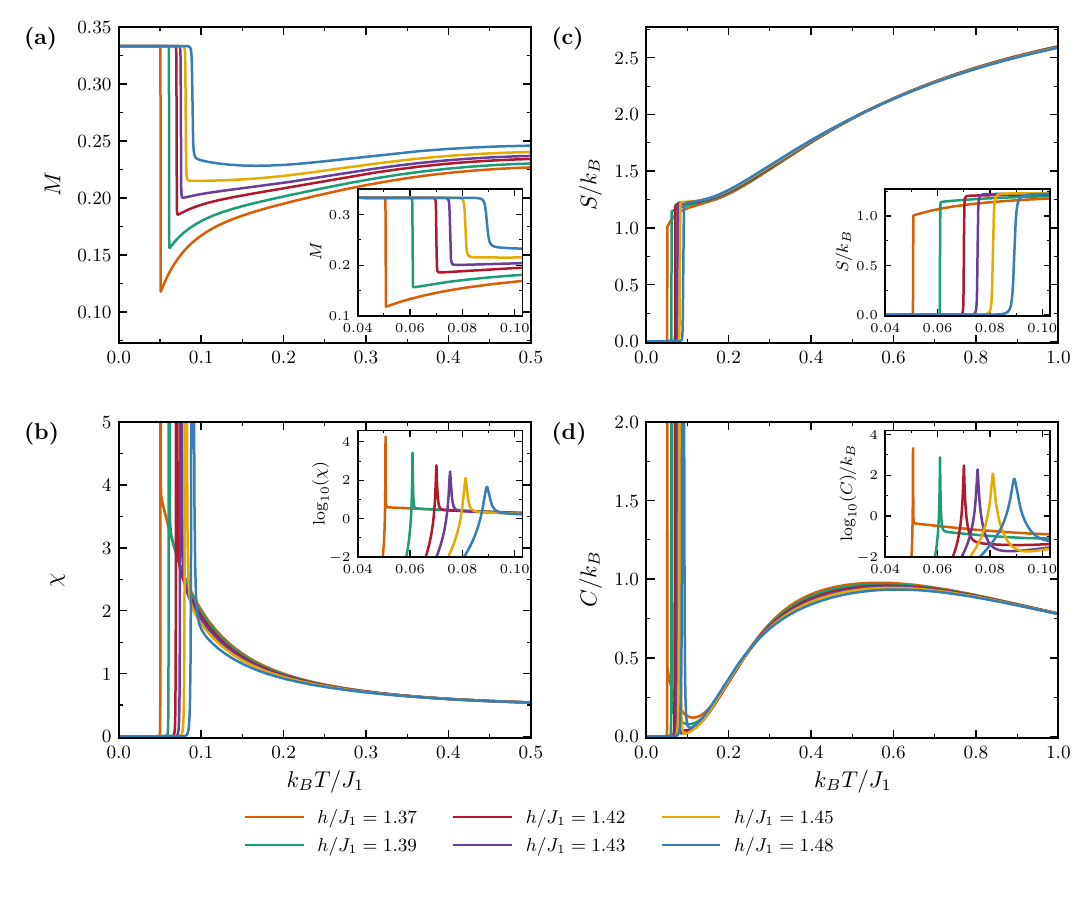}
	\end{center}
	\vspace{-0.5cm}
	\caption{Temperature dependence of the magnetization $M$ per spin (a), the susceptibility $\chi$ per spin (b), the entropy $S/k_{\rm B}$ per unit cell (c), and the specific heat $C/k_{\rm B}$ per unit cell (d) for the spin-1 Blume-Capel diamond chain with the interaction ratio $J_2/J_1=1.0$, the single-ion anisotropy $D/J_1=1.45$, and magnetic fields $h/J_1 \in [1.37, 1.48]$. Insets show the pseudo-transition region in an enhanced scale.}
	\label{fig4}
\end{figure}

Next, we consider the spin-1 Blume-Capel diamond chain with the coupling ratio $J_2/J_1=1.0$ and the single-ion anisotropy $D/J_1 = 1.45$. Fig.~\ref{fig4} illustrates evolution of the pseudo-transition phenomenon, which emerges under this condition as the system is driven into the vicinity of the triple point $[D/J_1; h/J_1] = [1.5; 1.5]$ where the nonmagnetic $[0,0,0]$, ferrimagnetic $[-1,+1,+1]$, and massively degenerate $[0,0,+1]$ ground states coexist. The magnetic and thermodynamic responses in this regime reflect this severe competition of phases through highly distinctive features. Fig.~\ref{fig4}(a) illustrates that the magnetization stays in a low-temperature regime at nearly constant value $M \approx 0.33$ consistent with the ferrimagnetic ground state $[-1,+1,+1]$ before exhibiting a sudden thermally-activated decrease to an intermediate value $M \approx 0.22$ subsequently followed by a less pronounced temperature dependence. The sharp thermally-induced decrease in the magnetization reflects the pseudo-transition from the ferrimagnetic phase $[-1,+1,+1]$ toward the nonmagnetic $[0,0,0]$ and massively degenerate $[0,0,+1]$ states. The significant redistribution of the dominant Boltzmann weights due to this pseudo-transition is also reflected in exceptionally sharp peaks of the magnetic susceptibility. Fig.~\ref{fig4}(b) clearly demonstrates that the magnetic susceptibility exhibits an extremely sharp finite peak at the pseudocritical temperature, whereas it does not display any pronounced high-temperature maximum unlike the previous case.  

The pseudo-transition is further corroborated by the temperature dependences of the entropy and specific heat shown in in Fig.~\ref{fig4}(c) and (d). The temperature dependencies of the entropy plotted in Fig.~\ref{fig4}(c) undergo a sudden step-like increase from zero to a relatively narrow intermediate plateau $S/k_{\rm B} \approx \ln (3) \approx 1.1$, which is successively followed by a gradual increase toward its extremal value. This abrupt yet continuous rise in the entropy is accompanied by a sharp anomaly in the specific heat, which is depicted in Fig.~\ref{fig4}(d) and its inset. Ultimately, the significant height of the low-temperature peak in the specific heat provides another distinct fingerprint of the pseudo-transition, which originates from a thermal activation of low-lying competing states massively concentrated around the triple coexistence point. 

\begin{figure}[t]
	\begin{center}
		\includegraphics[width=0.95\textwidth]{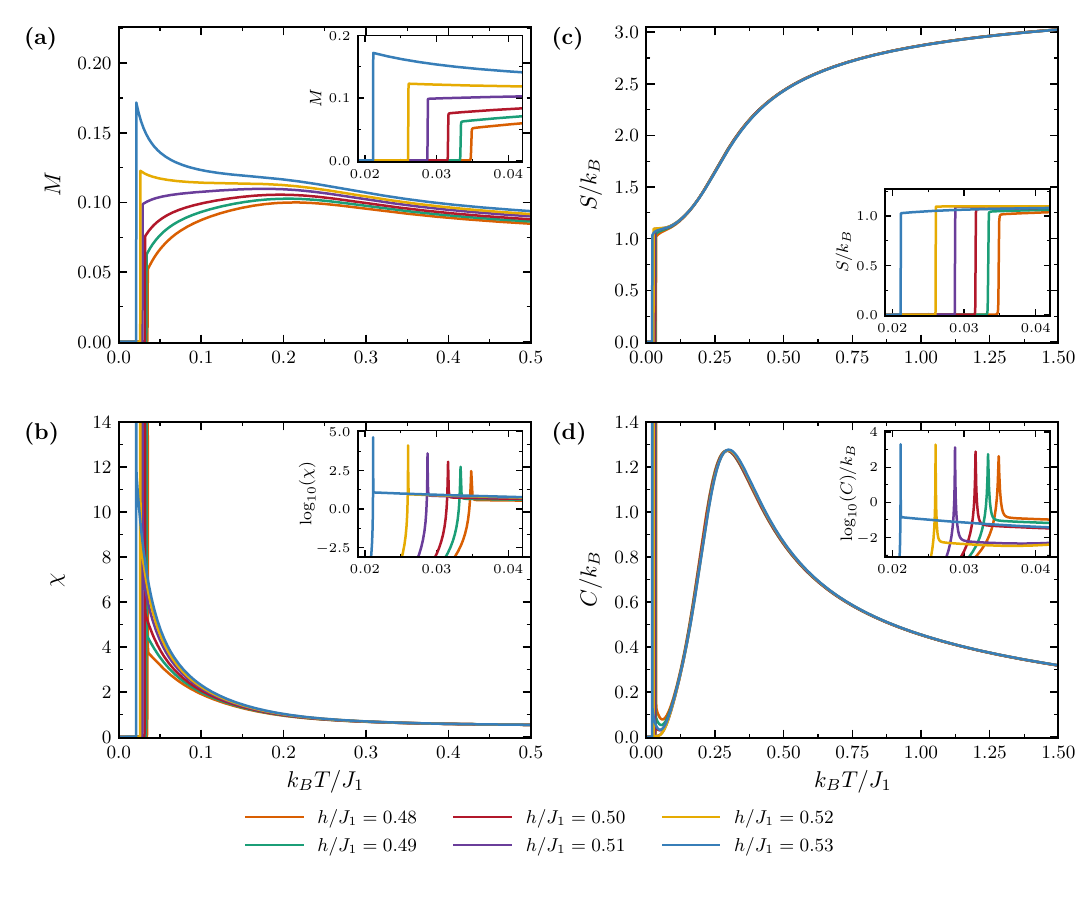}
	\end{center}
	\vspace{-0.5cm}
	\caption{Temperature dependence of the magnetization $M$ per spin (a), the susceptibility $\chi$ per spin (b), the entropy $S/k_{\rm B}$ per unit cell (c), and the specific heat $C/k_{\rm B}$ per unit cell (d) for the spin-1 Blume-Capel diamond chain with the interaction ratio $J_2/J_1=1.5$, the single-ion anisotropy $D/J_1=1.015$, and magnetic fields $h/J_1 \in [0.48, 0.53]$. Insets show the pseudo-transition region in an enhanced scale.}
	\label{fig5}
\end{figure}

As a final example, we consider the spin-1 Blume-Capel diamond chain with the interaction ratio $J_2/J_1 = 1.5$ and the single-ion anisotropy $D/J_1 = 1.015$. To achieve the pseudo-transition, the selected magnetic fields drive the investigated system close to the triple point $[D/J_1; h/J_1] = [1.0; 0.5]$, at which the nonmagnetic $[0,0,0]$, the ferrimagnetic $[-1,+1,+1]$, and the macroscopically degenerate antiferromagnetic $[-1,0,+1]$ phases coexist [see Fig.~\ref{fig2}(b)]. It is obvious from Fig.~\ref{fig5}(a) that the magnetization exhibits a zero plateau due to the nonmagnetic ground state $[0,0,0]$, which terminates at an abrupt thermally-induced rise evidencing the pseudo-transition. The temperature variations of the magnetic susceptibility shown in Fig.~\ref{fig5}(b) confirm the pseudo-transition through intense sharp peaks, which are subsequently followed by a relatively rapid temperature-induced decay observable in a high-temperature regime. 

The entropy versus temperature plot presented in Fig.~\ref{fig5}(c) likewise exhibits a steep but continuous increase at the pseudo-critical temperature. Upon further increase of temperature, the entropy displays a relatively narrow intermediate plateau $S/k_{\rm B} \approx \ln (3) \approx 1.1$ before gradually approaching its high-temperature limit. The steep low-temperature and gradual high-temperature increase in the entropy manifest itself also in a marked temperature dependence of the specific heat, which displays a sizable low-temperature anomaly well separated from a broad high-temperature maximum. This pronounced separation of energy scales confirms that the low-temperature anomaly originates from the entropy-driven pseudo-transition associated with thermally activated macroscopically degenerate low-lying states, whereas the broad high-temperature maximum reflects conventional thermal excitations.

\section{Conclusion}
\label{conclusion}

In this work, we have exactly investigated the ground state, magnetic and thermodynamic properties of the spin-1 Blume–Capel diamond chain in an external magnetic field by means of the transfer-matrix method. Although the present model belongs to one-dimensional spin systems for which finite-temperature phase transitions are strictly forbidden, we have demonstrated that the studied system exhibits remarkably sharp entropy-driven crossovers between a nondegenerate ground state and competing macroscopically degenerate low-lying excited states. These pseudo-transitions are reminiscent of finite-temperature phase transitions in that the magnetization and entropy display abrupt but continuous changes that closely mimic true discontinuities, while the magnetic susceptibility and specific heat exhibit extremely high yet finite peaks reminiscent of power-law divergences. These anomalous features associated with the pseudo-transitions are especially marked when the spin-1 Blume-Capel diamond chain is driven by control parameters such as an external magnetic field sufficiently close to a phase boundary or a triple point involving a macroscopically degenerate phase. The pseudo-transition can be thus viewed as an extremely sharp entropically-driven crossover, which arises from a thermal activation of massively degenerate low-energy excited states accumulated near the phase boundaries and triple points once the system is tuned to the nondegenerate ground state.  

The pseudo-transitions reported here for the spin-1 Blume-Capel diamond chain is not the unique feature of the present model, but it definitely represents a generic feature of more general class of one-dimensional spin systems. Although most of the one-dimensional systems displaying the pseudo-transition reported to date are either spin-1/2 or mixed-spin systems, it is quite plausible to conjecture that the spin-1 Blume–Emery-Griffiths diamond chain in a magnetic field represents another exactly solvable model exhibiting the pseudo-transition. Similar expectations are also valid for the spin-1 Ising–Heisenberg diamond chain \cite{ananikian2014, hovhannisyan2016}, which represents another platform for the exploration of the pseudo-transition physics. Beyond the magnetic structure of frustrated diamond chain, recent studies by Zvyagin and Zvyagina imply that the pseudo-transitions can be also found in other related frustrated one-dimensional spin chains such as the spin-1 Blume–Capel and Blume–Emery-Griffiths sawtooth chains \cite{zvyagin2026, zvyagin2026quasi}.

\section*{Acknowledgement}
The authors acknowledge funding by the EU NextGenerationEU through the Recovery and Resilience Plan for Slovakia under the project No. 09I03-03-V02-00021, the grant of Slovak Research and Development Agency under the contract No. APVV-24-0091, and the grant of The Ministry of Education, Research, Development and Youth of the Slovak Republic under the contract No. VEGA 1/0695/23.

\bibliographystyle{elsarticle-num} 
\bibliography{references}

\end{document}